\documentstyle[preprint,aps,epsf]{revtex}
\begin{document}
\preprint{
  \parbox{2.0in}{%
       hep-ph/9508267 \\
       MSU-HEP/50808 \\
       {\bf REVISED} \\
                     \\}
}
\title{Improving the Measurement of the Top Quark Mass}
\author{Jon Pumplin}
\address{
Department of Physics \& Astronomy \\
Michigan State University \\
East Lansing, MI \\
Internet address:  pumplin@msupa.msu.edu
}
\date{\today}
\maketitle
\begin{abstract}
Two possible ways to improve the mass resolution for observing
hadronic top quark decay $t \to b \, W \to 3 \, {\rm jets}$ are
studied:  (1) using fixed cones in the rest frames of the $t$
and $W$ to define the decay jets, instead of the traditional
cones in the rest frame of the detector; and (2) using the jet
angles in the top rest frame to measure $m_t / m_W$.  By Monte
Carlo simulation, the second method is found to give a useful
improvement in the mass resolution.  It can be combined with the
usual invariant mass method to get an even better mass
measurement.  The improved resolution can be used to make a more
accurate determination of the top quark mass, and to improve the
discrimination between $t \bar t$ events and background for
studies of the production mechanism.
\end{abstract}
\pacs{}
\narrowtext

\section {Introduction}
\label{sec:intro}

The discovery of the top quark in $p \bar p$ collisions \cite{cdf,d0}
was a milestone for the Standard Model.  The next steps beyond that
milestone will be to improve the measurement of the top mass, and to
test our understanding of QCD further by studying the details of its
production and decay.

We focus here on the ``single-lepton'' channel defined by
$p \bar p \to t \bar t X$ with one top decaying hadronically
($t \to W^+ \, b \to 3 \, {\rm jets}\,$, or its charge conjugate) and
the other decaying leptonically
($t \to W^+ \, b \to \ell^+ \, \nu_\ell + 1 \, {\rm jet}\,$, or its
charge conjugate, with $\ell = e$ or $\mu$).  This channel is
experimentally favorable because the high $p_\perp$ lepton and large
missing $p_\perp$ due to the neutrino provide a clean event signature
that naturally discriminates against backgrounds.  We will concentrate
on the hadronically decaying top, since the treatment of the
leptonically decaying one is complicated by errors in the measurement
of the transverse momentum of the neutrino, via missing $p_\perp$,
and by ambiguity in its longitudinal momentum.

We will not deal here with the question of how $t \bar t$ events are
to be identified \cite{cdf,d0,chao}.  Rather we will concentrate on
the problem of measuring $m_t$ once the events have been isolated.
The mass determination relies on measuring the energies and directions
of jets, and inferring from them the energies and directions of the
original quarks.  In addition to instrumental calibrations,
this requires corrections for the hard-scale branching of partons, and
for non-perturbative hadronization effects.  These hard and soft QCD
effects can be modeled by event generators such as
{\footnotesize HERWIG} \cite{herwig}, which have been shown to
describe the appropriate physics in $Z^0$ decay at the $e^+ e^-$
collider LEP \cite{lep,lepcone}, where the original $q \bar q$ energy
is precisely known.

Part of the top quark analysis involves measuring the mass of the
hadronically decaying $W$ boson.  An eventual goal of the analysis
should be to confirm the QCD effects in $W$ decay that are similar
to those seen in great detail for $Z^0$ decay.  In the meantime,
we can assume that the QCD effects are understood, and use the
$W$ mass for help in determining the correct jet assignments in
$t \bar t$ events.  The decay $W \to 2 \, {\rm jets}$ is also a
potentially useful tool to calibrate the detector, since the decay
jets probe the detector at a wide range of angles and jet transverse
momenta, and the $W$ mass is accurately known \cite{wmass}.  This
calibration is unavailable outside of top quark events, because
hadronic $W$ decays are generally obscured by large QCD
backgrounds \cite{howtoobserve}.

We will investigate two ideas that might improve the mass
resolution.  The first idea, examined in Sect.\ \ref{sec:cones}, is
to use fixed cones in the rest frames of the decaying $t$ and $W$
to define the jets, in place of the usual cones defined in the rest
frame of the detector.  The second idea, examined in
Sect.\ \ref{sec:angles}, is to use the angles between jets in
the $t$ rest frame to measure $m_t / m_W$ directly.

\section {Rest frame cone method}
\label{sec:cones}

The first idea that we will investigate can be understood most
clearly by considering the $W$ mass measurement.  Since the $W$
is color neutral, its decay products have nothing to do with other
jets present in the final state.  In the rest frame of the $W$,
the decays will look very similar to the $Z^0$ decays studied at
LEP.  They will almost always appear as back-to-back jets that can
be defined by two fixed cones of half-angle $\theta_0$ oriented in
opposite directions.  Reasonable cone sizes are on the order of
$\theta_0 = 45^\circ$.  Cones of this size are large enough to
contain a good fraction of the energies of the jets, which are
broadened by the collinear radiation singularity of QCD; while
they are small enough to leave a large fraction $\cos{\theta_0}$ of
the full $4 \pi$ of solid angle outside the two cones, so not too
much ``background'' is included in them.  Although cone algorithms
are not the traditional jet definitions in $e^+ e^-$ physics, they
have been shown to work there, and even to provide superior
resolution in jet angle and energy \cite{lepcone}.

This proposal to try fixed cones in the $W$ rest frame is quite
different from the current experimental procedure for top analysis,
where jets are defined by cones
$\sqrt{(\Delta \eta)^2+(\Delta\phi)^2} < R$ in the Lego variables
of the detector.  Those variables, $\eta = -\log\tan{\theta/2} =$
pseudorapidity and $\phi =$ azimuthal angle, are Lorentz invariant
only under boosts along the beam direction.  Meanwhile, the $W$ can
have a rather large component of boost transverse to the beam.  For
example, in $t \bar t$ events with typical acceptance cuts, the
middle $60\%$ of the $W$ transverse momentum distribution
extends from $40 \, {\rm GeV/c}$ to $110 \, {\rm GeV/c}$, so
$p_\perp^W$ is not small compared to $m_W$.  Lego-defined jet
cones therefore correspond to a wide variety of sizes and shapes
of ``cone'' in the $W$ rest frame, depending on the boost and
hence depending on details that have nothing to do with the $W$
decay.

Similarly, the jet cone for the $b$-quark jet from $t \to W \, b$
is most reasonably defined by a fixed angle in the $t$ rest
frame.  Since the $b$ carries color, the argument for a
cylindrically symmetric cone in this frame is weaker than in the
case of the $W$ decay.  But the $t$ rest frame is nevertheless at
least more logical than using the rest frame of the detector.

It would similarly be reasonable to define the jet cone for the
other $b$-jet, from the semileptonically decaying top, in the
rest frame of that top.  That should be done when this method is
applied to real data.  But we ignore it for the moment, in order
to avoid bringing in the issue of how to estimate the longitudinal
momentum of the neutrino, which is needed to find that frame.

The idea of using fixed cones in the decay rest frames to define
the jets is easy to implement as a follow-on to the traditional
top quark analysis with ``kinematic reconstruction'' \cite{cdf},
{\it i.e.}, with explicit matching of the observed jets to decay
partons.  The procedure for each event is:
\begin{enumerate}
\item
Identify the four-momenta of the jets corresponding to
$t \to j_1 \, j_2 \, j_3$ and $t \to j_4 \ell \nu$, where $j_1$,
$j_2$ come from $W$ decay and $j_3$, $j_4$ are $b$-quark jets, by
the usual procedures of CDF or D\O.  $\,b$-jet tagging is helpful
for this, but not essential.  There may of course be additional
jets observed in the event.

\item
Define $j_F$ and $j_B$ to be four-momenta directed in the forward
and backward beam directions:  $j_F = (1;0,0,1)$ and
$j_B = (1;0,0,-1)$ for beams in the $z$-direction.

\item
Redefine the 4-momentum of $j_1$ as the sum of the 4-momenta
observed in the calorimeter detector in all cells (``towers'')
that lie within $\theta_0$ of the old $j_1$ in the rest frame of
the old $j_1 + j_2$.  However, do not include cells that lie closer
in angle to $j_2$, $j_3$, $j_4$, $j_F$, or $j_B$ than to $j_1$.

Similarly redefine the 4-momentum of $j_2$ as the sum of the
4-momenta of all cells that lie within $\theta_0$ of the old $j_2$
in the rest frame of the old $j_1 + j_2$, and are not closer in
angle to $j_1$, $j_3$, $j_4$, $j_F$, or $j_B$.

Similarly redefine the 4-momentum of $j_3$ as the sum of the
4-momenta in cells that lie within $\theta_0$ of the old $j_3$ in
the rest frame of the old $j_1 + j_2 + j_3$, and are not closer
in angle to $j_1$, $j_2$, $j_4$, $j_F$, or $j_B$.

\item
Replace the old jet 4-momentum estimates by the new ones and
repeat Step 3 a few times.  This iteration converges completely
for $> \! 99.8\%$ of events.  A maximum of 10 iterations is
sufficient, since the momenta stop changing before that for
$> \! 99.5\%$ of events.

\end{enumerate}

The major new work is in Step 3, which is very easy to implement
in the following way.  Treat the energy in every cell of the
detector that receives energy above its noise level as if it
came from a zero-mass particle.  There will typically be a few hundred
such cells in each event.  Make a list of their four-momenta.
The appropriate elements of this list to be added in the various
parts of Step 3 can be found without making explicit Lorentz
transformations, using the exact formula for the angle $\theta$
between $\vec{p}$ and $\vec{q}$ in the rest frame of $r$
\begin{eqnarray}
\cos{\theta} =
\frac
{p \! \cdot \! r \, q \! \cdot \! r - p \! \cdot \! q \, r^2}
{\sqrt{[(p \! \cdot \! r)^2 - p^2 \, r^2] \,
[(q \! \cdot \! r)^2 - q^2 \, r^2]}}
\; ,
\label{eq:costhe}
\end{eqnarray}
which holds for any four-vectors $p$, $q$, $r$.

To test the idea, {\footnotesize HERWIG 5.7} \cite{herwig} was used
to simulate the single-leptonic top channel at the Tevatron energy
$\sqrt{s} = 1.8 \, {\rm TeV}$.  Typical experimental cuts were
approximated by
$|\eta^\ell| < 2.0 \,$,
$p_\perp^\ell > 20 \, {\rm GeV/c} \,$,
$p_\perp^\nu > 25 \, {\rm GeV/c} \,$.
The four hadron jets were required, at the partonic level, to have
$p_\perp^{\rm \, jet} > 20 \, {\rm GeV/c}$ and to be isolated in
Lego from the lepton by
$\sqrt{(\Delta \eta)^2+(\Delta\phi)^2} > 0.5$ and from each other
by $\sqrt{(\Delta \eta)^2+(\Delta\phi)^2} > 0.8\,$.  The detector
was simulated by an array of $0.1 \times 0.1$ cells in
$(\eta, \phi)$, covering the region $-4.5 < \eta < 4.5 \,$.
Gaussian energy resolutions with $\Delta E / E = 0.55 / \sqrt{E}$
for charged hadrons and $\Delta E / E = 0.15 / \sqrt{E}$ for
leptons and photons were assumed (in ${\rm GeV} = 1$ units).
A small amount of shower spreading was also included:  the energy
of each final hadron was spread over a disk of radius $0.1$
in the $(\eta, \phi)$ plane, before being deposited in the
appropriate calorimeter cells.

A cone-type jet finding algorithm described
previously \cite{howtotell} was used to search for jets in the
simulated calorimeter.  A small cone size $R = 0.3$ was used,
as is standard practice in top quark analysis (CDF uses $R = 0.4$)
to reduce the loss of events from overlapping jets, since one is
asking for 4 jets within a rather small area in Lego --- {\it e.g.},
all four decay jets lie within some strip of width $2.0$ in $\eta$
in $62\%$ of the events.  A minimum observed jet
$p_\perp$ of $15 \, {\rm GeV/c}$ was required.  A feature of my
jet-finder \cite{howtotell} is that it finishes with an iteration
in which each cell of the detector that lies within $R$ of
at least one jet axis is assigned to the nearest jet axis.  The
jet momenta are then recomputed and this procedure is repeated.
This feature tends to improve the mass resolution for objects
decaying into jets.  It is similar to the procedure being tried
here, except that the nearest jets are now to be defined by angles
in the appropriate rest frames.

To obtain a sample that could be cleanly interpreted, the
``observed'' jets in the simulation were matched to their original
quark partons, using a criterion based on the agreement in Lego
variables $\eta$ and $\phi$, along with a contribution from the
agreement in $p_\perp$.  A good match was obtained for $85\%$ of
the events, upon which Fig.\ 1 and Fig.\ 2 are based.

Fig.\ 1 shows the histogram of the mass observed for
$W \to {\rm jj}$.  The {\it dotted curve} is for jets defined
by $\sqrt{(\Delta \eta)^2+(\Delta\phi)^2} < R = 0.3$
in the Lego variables of the detector.  Note that the center of
the peak is well below the actual $m_W = 82 \, {\rm GeV/c^2}$
assumed in the simulation, and that the peak is very asymmetrical,
with a tail toward lower ${\rm jj}$ masses.  This skewing of the
mass distribution toward low mass is a direct result of gluon
radiation falling outside the cone, which is characteristic of QCD,
and which has been verified at LEP.  It demonstrates that making
QCD corrections is a very important part of the top quark mass
measurement, especially if a cone size as small as $0.3$ is used.

The {\it solid curve} in Fig.\ 1 shows the result of redefining
the jets using a larger cone size $R = 0.7 \,$.  This included an
iteration in which each calorimeter cell that lies within $R$ of
at least one of the 3 jet axes from hadronic $t$ decay is
reassigned to the jet with the nearest of those axes.  This
iteration is crucial to obtaining the improvement in mass
resolution shown.  Note that the qualitative QCD feature of
skewing toward lower mass is still visible, but much less
pronounced.  Also, the center of the peak is at a higher mass,
closer to the true partonic mass.

The cone size $R = 0.7$ is about optimal.  If the width of the mass
peak is defined by the range $\Delta M$ that contains the middle
$50\%$ of the probability distribution, then $\Delta M / M$
goes from $0.197$ to $0.151$ to $0.133$ as $R$ is increased from
$0.3$ to $0.5$ to $0.7\,$.

The {\it dashed curve} in Fig.\ 1 shows the result of redefining the
jets in the appropriate decay rest frames by the steps enumerated
above.  The curve shown is for $\theta_0 = 45^\circ$, which is about
optimal.  One sees that the mass resolution obtained this way is
quite good ($\Delta M / M = 0.138\,$), but not as good as the
$R = 0.7$ curve.\footnote{
An early version of this paper claimed that defining jet cones in
these ``appropriate'' rest frames led to superior mass resolution.
The improvement found actually resulted from the fact that the
rest-frame cones were on average larger than the Lego cones used
there, in concert with the iterations in which each calorimeter
cell is associated with the nearest jet axis.  As long as such
iterations are included, conventional Lego cones of size
$R \simeq 0.7$ provide as good or slightly better resolution,
as shown here.}

Fig.\ 2 shows a similar histogram of the mass observed for
$t \to W \, b \to {\rm jjj} \,$.  One again sees QCD skewing
of the peak toward lower masses, relative to the value
$m_t = 175 \, {\rm GeV/c^2}$ assumed in the simulation.
This mass shift includes a contribution of
$\, \approx -4 \, {\rm GeV/c^2}$ from losses in observed
jet energy due to neutrinos.  One again sees a dramatic
reduction in skewing and improvement in the resolution, when
the cone size is increased from $R=0.3$ ({\it dotted curve})
to $R=0.7$ ({\it solid curve}).  We emphasize that to obtain
this improvement, it is essential to use the iterative
procedure whereby each calorimeter cell is assigned to the
nearest decay jet axis.

The cone size $R = 0.7$ is again about optimal. Defining as
before the width of the peak by the range $\Delta M$ that
contains the middle $50\%$ of the probability distribution,
$\Delta M / M$ goes from
$0.186$ to $0.149$ to $0.125$ as $R$ is increased from
$0.3$ to $0.5$ to $0.7\,$.

The suggestion to define the jets using a fixed cone angle
in the rest frame of the $t$ for the $b$-jet, and in the rest
frame of the $W$ for its decay jets, which was proposed in
this section, leads to the {\it dashed curve} in Fig.\ 2.
One sees that, as in Fig.\ 1, this plausible idea does not
in fact improve the mass resolution ($\Delta M / M = 0.130$),
compared to $R=0.7$ cones --- although it is not much worse,
either.

\section {Jet angle method}
\label{sec:angles}

So far, we have computed the observed masses of $t$ and $W$
directly from the 4-momenta of the jets into which they decay,
as simply the invariant mass of the sum of those jet momenta.
The estimated 4-momentum of each jet is obtained by adding the
4-momenta detected by the calorimeter cells inside its jet cone.
The $t$ and $W$ are thus formally treated as decaying into a
large number of zero-mass particles that represent the energies
deposited in the detector.  Defined in this way, the individual
jets have sizeable invariant masses --- often on the order of
$10 \, {\rm GeV/c^2}$ or higher.

A possible alternative method to determine the top mass in each
event would be to use only the {\it directions} of the jets in
the top rest frame, treating the jets as zero-mass objects.  To
do this, let $\psi_{1}$ and $\psi_{2}$ be the angles between the
$b$-jet and the two jets from $W$ decay in that frame.  Then
$\psi_{3} = 2\pi - \psi_{1} - \psi_{2}$ is the angle between
the $W$ decay jets.  According to zero-mass kinematics,
\begin{eqnarray}
\frac{m_t}{m_W} =
\frac
{\sin{\psi_{1}} \, + \, \sin{\psi_{2}} \, + \, \sin{\psi_{3}}}
{\sqrt{2 \, \sin{\psi_{1}} \, \sin{\psi_{2}} \,
(1 - \cos{\psi_{3}})}}
\; .
\label{eq:angles}
\end{eqnarray}
The angles needed in this formula can be obtained from the measured
jet 4-momenta in any frame using Eq.~(\ref{eq:costhe}).  At the
partonic level, the use of zero-mass kinematics is promising since,
{\it e.g.}, the energy of the $b$-quark is
$\approx 70 \, {\rm GeV}$, so the influence of its mass
$m_b \approx 5 \, {\rm GeV/c^2}$ is very small.

An advantage of this technique is that it measures the top quark
mass in units of $m_W$, which is independently known \cite{wmass}.
Because a dimensionless ratio is measured, the result for $m_t$
is insensitive to any constant overall scale factor in the energy
measurements.

For comparison with the results of Sect.\ \ref{sec:cones}, the
mass distribution obtained using this jet angle method are plotted
as the {\it dot-dash curve} in Fig.\ 2, by scaling with the value
of $m_W$ assumed in the simulation.  {\it The mass distribution
is narrower than the best previous result by about $6\%$.}
In addition to improved mass resolution, the curve is seen
to be nearly symmetrical, and the center of the peak is very close
to the mass value $175 \, {\rm GeV/c^2}$ assumed in the simulation.
The shift in the peak value due to the non-detection of neutrinos
in the jets is also somewhat smaller when the jet angle method is
used.

When the simulation is repeated using
$m_t = 190 \, {\rm GeV/c^2}$, the center of the peak in
$m_{\rm jjj}$ determined by the jet angle method again comes
out within $1 \, {\rm GeV}$ of the input $m_t\,$, so the good
features found above are not specific to the assumed mass value.
The nearly exact agreement with the input mass is of course
partly accidental --- {\it e.g.}, the output mass would shift
upward by $\sim \! 3 \, {\rm GeV}$ if the calorimeter were able
to detect neutrinos.  The important result is that the output
mass is close to the input one, and tracks it proportionally.

The curve shown is for jets defined by $R=0.7$, which is
approximately optimal; but a further advantage of the jet angle
method is that it is found to be less sensitive to the choice of
$R$ than is the traditional invariant mass method.  Defining the
jets using fixed cones in the rest frames as described in
Sect.\ \ref{sec:cones}, and then using the angle method to
compute $m_t/m_W$ also works, but not quite as well as simply
using the Lego definition with $R=0.7\,$.

\section {Direct test}
\label{sec:direct}

So far, we have explicitly used the original parton momenta,
which are known in the simulation program, to determine the
correct assignments of the observed jets.  In real life, of
course, that luxury is unavailable, so one faces a combinatoric
problem in analyzing semileptonic $t \bar t$ events.  There
are 12 ways to assign 4 observed jets to the $b$-jet from
semileptonic $t$ decay, the $b$-jet from hadronic $t$ decay,
and the two jets from hadronic $W$ decay.  There may be fewer
combinations to try if one or both of the $b$-jets has been
tagged; but there may be many more, since frequently there are
additional jets, {\it e.g.}, from initial state radiation.
Hence mass histograms like Figs.\ 1 and 2 will be contaminated
by a background from events in which the jets are misassigned.

In order to test our methods directly, the {\footnotesize HERWIG}
events and calorimeter simulation described above were used
without making parton level cuts, and without using any of the
partonic information to determine the correct jet assignments,
in the following simplified event analysis:
\begin{enumerate}
\item
Jets with observed $E_\perp > 15 \, {\rm GeV}$ are found using
$R=0.4$ cones in $(\eta, \phi)$ space.  If fewer than 4 jets
are found, the event is rejected.

\item
If more than 4 jets are found, only those with the 4 largest
values of $p_\perp$ are kept.  We ignore $b$-tagging here, so
all 12 ways of assigning the jets to the 4 expected jets are
examined.  (By trying all possible subsets of 4 jets, it would
be possible to reclaim an additional $\sim \! 10\%$ of good
events; but the combinatoric and non-top backgrounds to those
events would be considerably larger, and the mass resolution
not quite as good, so this will be of doubtful value even in
the final data analysis.)

\item
An iteration similar to that described in
Sect.\ {\ref{sec:cones}} is carried out, so that each
calorimeter cell within $R=0.7$ of any one of the 4 significant
jet axes is associated with the nearest of those axes.

\item
The transverse momentum of the neutrino is estimated as usual
as the negative of the total observed $\vec{p}_\perp$.  Its
longitudinal momentum is estimated by requiring $m_{\ell \nu}$
to equal the $W$ mass used in the simulation.  The magnitude
of the rapidity difference $|\eta_\nu - \eta_\ell|$ is then
given by
\begin{eqnarray}
m_{\ell \nu}^2 = 2 \, p_\perp^{\ell} \, p_\perp^{\nu} \,
[\cosh(\eta_\nu - \eta_\ell) - \cos(\phi_\ell - \phi_\nu)] \; .
\label{eq:massellnu}
\end{eqnarray}
The sign of $\eta_\ell - \eta_\nu$ is assumed to be the same
as the sign of $\eta_\ell$, which corresponds to choosing
between the two possible solutions for $p_\parallel^\nu$ the
one giving the smaller energy for the leptonically decaying
$W$.  If Eq.~(\ref{eq:massellnu}) has no solution (because
$\cosh(\eta_\nu - \eta_\ell) < 1$), we take
$\eta_\nu = \eta_\ell$ \cite{chao}; or reject the event
if there is no solution with $m_{\ell \nu} < 100$.

\item
We reject some candidate jet assignments immediately on the basis
of very loose cuts:  we require $120 < m_{b \ell \nu} < 220$ for
the leptonically decaying top and $55 < m_{\rm jj} < 100$ for the
hadronically decaying $W$.  (Broad cuts on the hadronic top mass,
or on the minimum non-$W$ dijet mass in the hadronic top might
also be useful at this point \cite{chao}.)

\item
To select the correct jet assignment according to the best fit
to the leptonic top and hadronic $W$ masses, we define a measure
of fit by
\begin{eqnarray}
D   & = & \sqrt{D_1^{\,2} + D_2^{\,2}} \nonumber \\
D_1 & = & (m_{b \ell \nu}^3 - m_1^3)/
(3 \, m_1^2 \, \Delta_1) \nonumber \\
D_2 & = & (m_{jj}^2 - m_2^2)/
(2 \, m_2 \, \Delta_2) \; .
\label{eq:D1D2}
\end{eqnarray}
The parameters $m_1 = 169$, $\Delta_1 = 16$, $m_2 = 78.3$,
$\Delta_2 = 6.7$ in ${\rm GeV} = 1$ units were found using
the previous simulation, by making the range
$-1.0 < D_1 < +1.0$ correspond to the middle $60\%$ of
the probability distribution for $m_{b \ell \nu}$ for the
correct jet assignment, while $-1.0 < D_2 < +1.0$ corresponds
similarly to the middle $60\%$ of the $m_{\rm jj}$
distribution for $W$ decay.  The cubic ($m_{b \ell \nu}^3$)
and quadratic ($m_{jj}^2$) forms in Eq.~(\ref{eq:D1D2})
were chosen because these quantities have more symmetrical
probability distributions than the QCD-skewed masses
themselves.  The factors in the denominators were chosen to
make the constants $\Delta_1$ and $\Delta_2$ correspond
directly to the widths of the distributions.

When it comes time to analyze real data, the parameters in
Eq.~(\ref{eq:D1D2}) can be tuned using Monte Carlo events.
Also, the parameters in $D_1$ should be adjusted to be
self-consistent with $m_t$ as measured by the hadronic $t$
decays.  Actually, the analysis could no doubt be improved
by allowing for differences in the expected range for the
observed $m_{b \ell \nu}$, based on the configuration of
$\ell $ and $\nu$ momenta.  Additional contributions to
the measure of fit for each possible jet assignment could
also be included, {\it e.g.}, based upon the transverse
and longitudinal momenta of the $t$ and $\bar t$.

\item
The jet assignment that gives the lowest value of $D$ is
assumed to be the correct one.  A small number of events
($8\%$) with a poor fit ($D > 2.5$) are rejected.   A further
$17\%$ of events are rejected when any two of the four final
jet axes are separated by less than $0.7$ in Lego.  Otherwise,
the corresponding mass for $t \to {\rm jjj}$ is entered into
a histogram to create Fig.\ 3.

\end{enumerate}

Note that we adopt the idea advocated in Ref.\ \cite{chao}
that the $t$ mass measurement should be based on the
hadronically decaying top.  The mass of the leptonically
decaying one contributes only to choosing the correct set
of jet assignments.  This avoids a number of possibilities
for systematic errors due to uncertainties in the neutrino
momentum measurement.  Only the central value assumed for
the leptonic decay mass needs to be tuned to match the final
average $m_t$ found in the experiment.  If the assumed value
is incorrect, it will reduce the signal rate, but should
not change the measured hadronic top mass.

Fig.\ 3 shows the $m_{\rm jjj}$ distributions for
$t \to b W \to {\rm jjj}$.  The {\it dotted curve} shows
the invariant mass given by the
original $R=0.4$ cones.  The {\it solid curve} shows the
invariant mass given by $R=0.7$ cones, including the
iteration that assigns calorimeter cells to the nearest
decay jet axis.  Note that going from $R=0.4$ to $R=0.7$
yields a dramatic improvement in the sharpness of the peak,
and considerably reduces the downward shift in mass relative
to the $m_t = 175$ input to the simulation.  The benefits of
going to $R=0.7$ cones is actually even greater than what is
shown here, because the larger cones were used to infer the
correct jet assignments for both curves.

The {\it dot-dashed curve} shows the result of calculating
$m_{\rm jjj}$ by the jet angle method of
Sect.\ {\ref{sec:angles}}.  It shows a clear peak that is more
symmetrical and is centered closer to the input value of $m_t$,
confirming the results of Sect.\ {\ref{sec:angles}}.  The peak
is not quite as high as the $R=0.7$ result, which may be due
to the fact that the ordinary cone method is somewhat less
affected by misassigned jets, since the value it gives for
$m_{\rm jjj}$ is independent of which of the three dijet pairs
is identified as the hadronically decaying $W$.

The jet angle method and the ordinary invariant mass offer
two largely independent measurements of $m_{\rm jjj}$ in
each event.  We can combine the two measurements to obtain
an average that has smaller errors than either measurement by
itself.  To allow for the fact that the two measurements contain
different systematic mass shifts, it is seems appropriate to
combine them by taking their geometric mean
$m_{\rm ave} = \sqrt{m_{\rm jjj}^{(1)} \, m_{\rm jjj}^{(2)} }$,
although the distribution of their simple average is almost
identical.  The resulting mass histogram, shown as the
{\it dashed curve} in Fig.\ 3, displays the best peak of all.
This mass variable is therefore most promising for future
analyses of $t \bar t$ data.

\section {Conclusions}
\label{sec:conclusion}

In summary, we have seen that
\begin{enumerate}
\item
To obtain the best mass resolution for hadronic top decays,
it is essential to use a sufficiently large cone size
$R \simeq 0.7$ in the Lego variables $(\eta$, $\phi)$ of the
detector, and to include an iteration whereby final particles
within each of the four jet cones
$\sqrt{(\Delta \eta)^2+(\Delta\phi)^2} < R$ of the
$t \bar t$ final state are self-consistently assigned to the
nearest axis.  Quantitatively, going from $R = 0.3$ to $R = 0.5$
reduces the width of the top mass peak by a factor
$\approx \! 0.80\,$.  Going on from $R = 0.5$ to $R = 0.7$
reduces it by a further factor $\approx \! 0.86\,$.

\item
A suggestion to define the jets by cones of fixed angle in
the rest frames of the decaying $t$ and $W$ was argued to be
plausible and shown to be simple to carry out, but did not in
fact improve the mass resolution.  Instead, it gave results
similar to $R=0.7\,$.

\item
A suggestion to compute the top mass in each event on the basis
of the jet angles in the measured $t$ rest frame, using the
zero-mass kinematical formula Eq.~(\ref{eq:angles}), was shown
to make a significant improvement in the measurement of $m_t$.
It reduces the width of the top mass peak slightly, and makes
the peak more symmetrical and less sensitive to the choice of
$R$.  This jet angle method is very easy to apply, since the
necessary angles are easily computed from the jet four-momentum
estimates using Eq.~(\ref{eq:costhe}).  A further advantage of
this method is that it measures $m_t$ as a ratio to the known
$m_W$, and is thus insensitive to the absolute calibration for
jet energy measurements.

\item
An even better measure of $m_t$ in each event can be obtained
using the geometric mean
$m_{\rm ave} = \sqrt{m_{\rm jjj}^{(1)} \, m_{\rm jjj}^{(2)}}$
of the two measures of $m_{\rm jjj}$ given by the ordinary
Lorentz-invariant mass of the three-jet system, and the jet
angle method.

\end{enumerate}
These conclusions have been confirmed by the simulation of
Sect.\ {\ref{sec:direct}}, which is a simplified version
of the essence of actual $t \bar t$ data analysis.

To clarify the point regarding cone size, let us examine the
standard experimental practice of ``correcting'' the jet
$p_\perp$ measured in a small cone, by a factor that is
determined from Monte Carlo simulation, or experimentally,
{\it e.g.}, by studying the $p_\perp$ balance of hard
scattering in events with a direct photon and a jet.  This
correction factor could in general be a function of $p_\perp$
and $\eta$.  However, no such ``out-of-cone correction'' can
improve the trijet mass resolution on a par with actually
{\it using} larger cones, since when the latter is done, the
revisions in $p_\perp$ are found to vary widely from
case to case as a direct result of the QCD collinear
radiation singularity.  This is shown in Fig.\ 4, which
displays the distribution of jet transverse energy increases
in going from $R=0.4$ to $R=0.7$ in the simulation of
Sect.\ {\ref{sec:direct}}, for jets from events in the signal
peak with a typical $p_\perp \approx 40$ in the smaller cones.
Quantitatively, the middle $68\%$ of the probability
distribution extends from $1.0$ to $6.7$ for the $b$-jets,
and from $0.8$ to $5.8$ for the jets from $W$ decay.

The importance of ample cone size, with final state energies
assigned to the nearest cone, has been clearly demonstrated.
The jet angle method has been shown to have significant
advantages.  Although the rest-frame cone method was found
to be no better than the standard procedure, it should still
be pursued on the grounds that its systematic errors might
be different from the conventional analysis.

Most importantly, the mass variable $m_{\rm ave}$, defined
as the geometric mean of the two estimates of $m_t$ using
$R=0.7$ cones, should be tried since it produced the best
mass resolution of all methods examined here.  The improved
resolution should help to reduce the statistical uncertainty
in the location of the mass peak, which will arise from the
rather small number of events ($\sim \! 100$) expected
in each experiment by the end of the current Tevatron running
period.  This technique should therefore improve the accuracy
of the top quark mass measurement.

The next steps, which can only be carried out by the
experimenters of CDF and D\O, must be to try these procedures
using full detector simulations, and then real data.

The improvements in mass resolution could also help with the
the difficult problem of observing doubly hadronic top events
$t \bar t \to (Wb) \, (W \,b) \to 6 \, {\rm jets}$ \cite{giele}.

\section*{Acknowledgments}
I thank S. Kuhlmann, J. Huston, J. Linnemann, T. Ferbel, and
S. Snyder for information about the experiments.  This work was
supported in part by U.S. National Science Foundation grant
number PHY-9507683.

\newpage

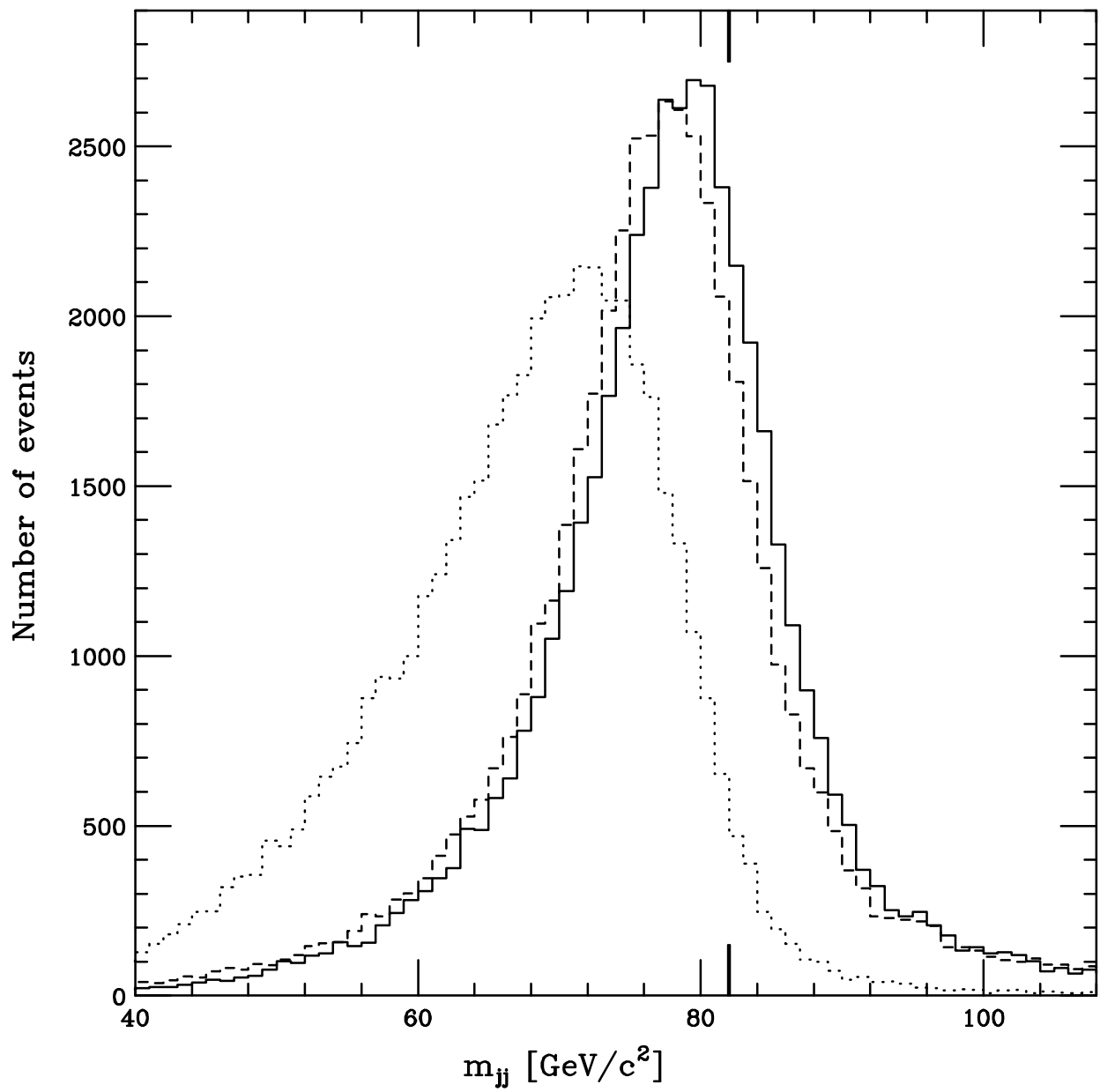
\begin{figure}
\caption{\protect
Simulated dijet invariant mass distribution for
$\protect W \to {\rm jj}$ from semileptonic top events.
The {\it dotted} curve is for jets defined by
$\protect \sqrt{(\Delta \eta)^2+(\Delta\phi)^2} < 0.3$
cones in the rest frame of the detector.
The {\it solid} curve is for jets defined by
$\protect \sqrt{(\Delta \eta)^2+(\Delta\phi)^2} < 0.7$
cones in the rest frame of the detector.
The {\it dashed} curve is for jets defined by
$\protect \theta < 45^\circ$ cones in the $W$ rest frame,
via the iteration proposed in Sect.\ {\protect \ref{sec:cones}}.
A heavy tick mark indicates the value of $\protect m_W$ assumed
in the simulation.
}
\label{figure1}
\end{figure}

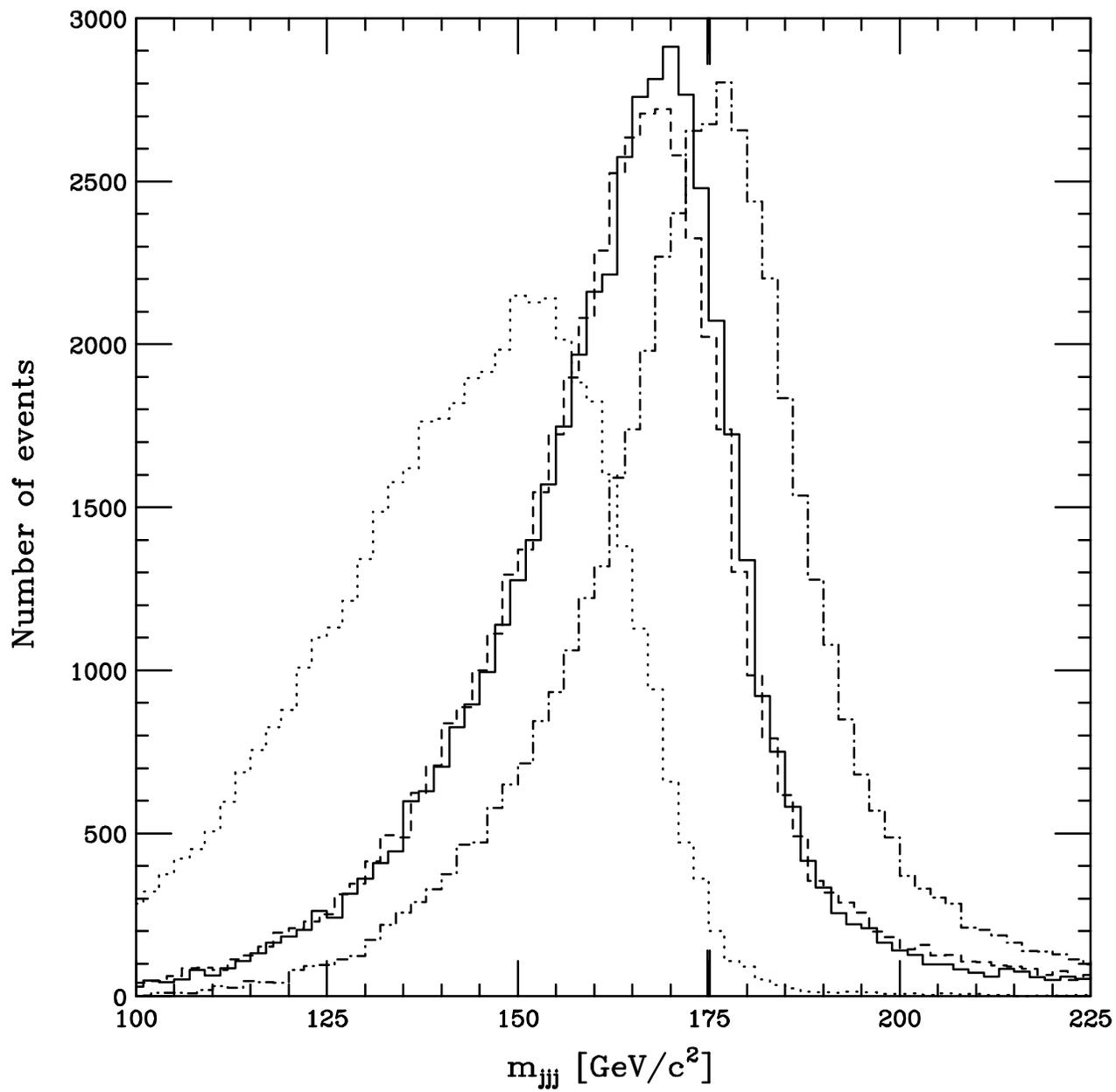
\begin{figure}
\caption{
Simulated trijet invariant mass distribution for
$\protect t \to W \, b \to {\rm jjj}$ from semileptonic top
events.  As in Fig.\ 1,
the {\it dotted} curve is for jets defined by
$\protect \sqrt{(\Delta \eta)^2+(\Delta\phi)^2} < 0.3$ cones,
the {\it solid} curve is for
$\protect \sqrt{(\Delta \eta)^2+(\Delta\phi)^2} < 0.7$ cones, and
the {\it dashed} curve is for
$\protect \theta < 45^\circ$ cones
in the $t$ rest frame for the $b$-jet and in the $W$ rest
frame for the other two.
The {\it dot-dashed} curve is for
$\protect \sqrt{(\Delta \eta)^2+(\Delta\phi)^2} < 0.7$ cones
with $m_{\rm jjj}$ computed by the ``jet angle method'' of
Sect.\ {\protect \ref{sec:angles}}.  A heavy tick mark indicates
the value of $\protect m_t$ assumed in the simulation.
}
\label{figure2}
\end{figure}

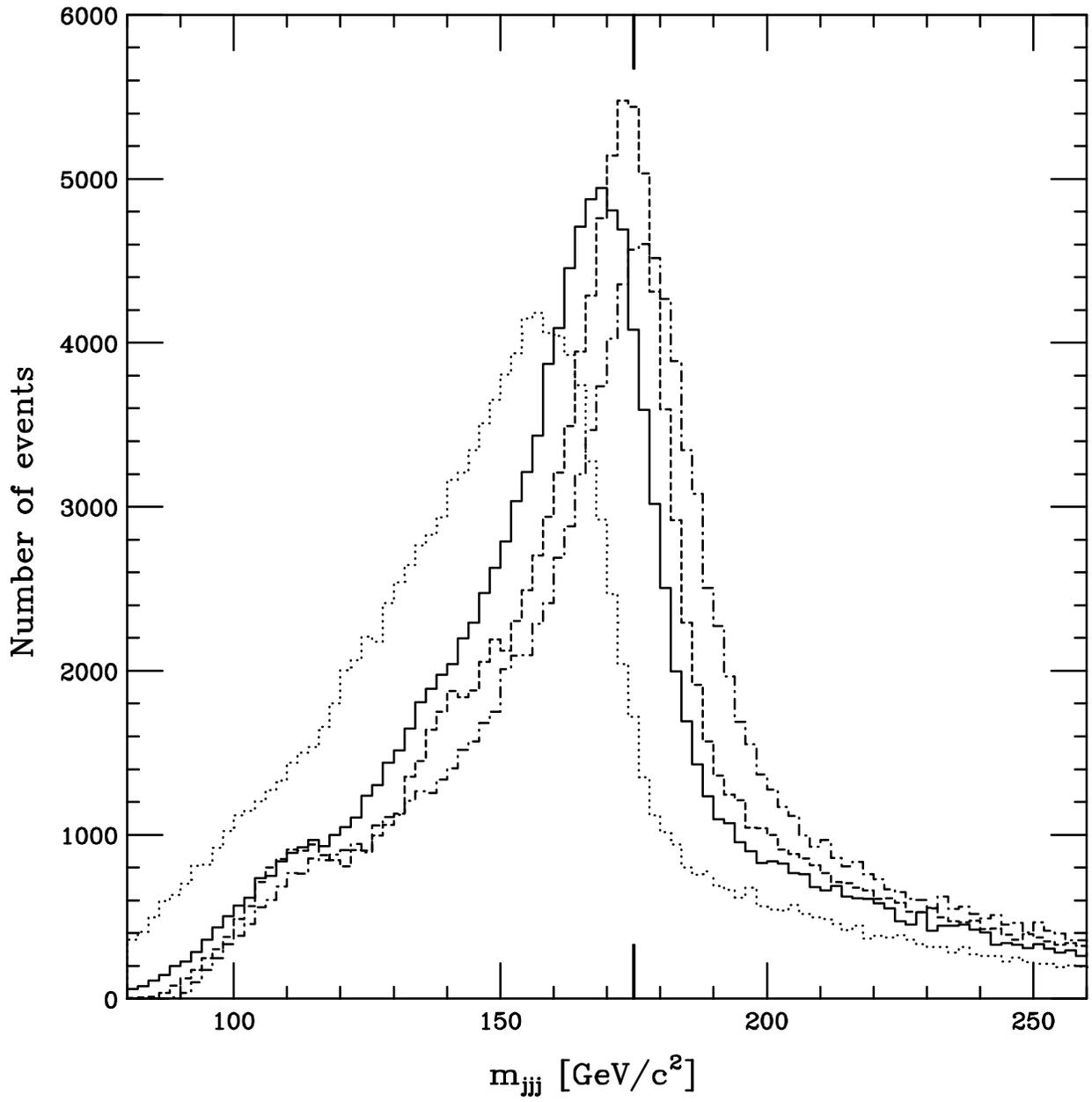
\begin{figure}
\caption{\protect
Trijet invariant mass distribution for
$\protect t \to Wb \to {\rm jjj}$ from the more complete
simulation of Sect.\ {\protect \ref{sec:direct}}.
The {\it dotted} curve is for jets defined by $R = 0.4$ cones.
The {\it solid} curve is for jets defined by $R = 0.7$ cones.
The {\it dot-dashed} curve is for the same $R = 0.7$ cones,
with $m_{\rm jjj}$ computed by the ``jet angle method'' of
Sect.\ {\protect \ref{sec:angles}}.
The {\it dashed} curve, which gives the best resolution of all,
is for the same $R = 0.7$ cones, with $m_{\rm jjj}$ computed as
the geometric mean of the two previous mass measures.
}
\label{figure3}
\end{figure}

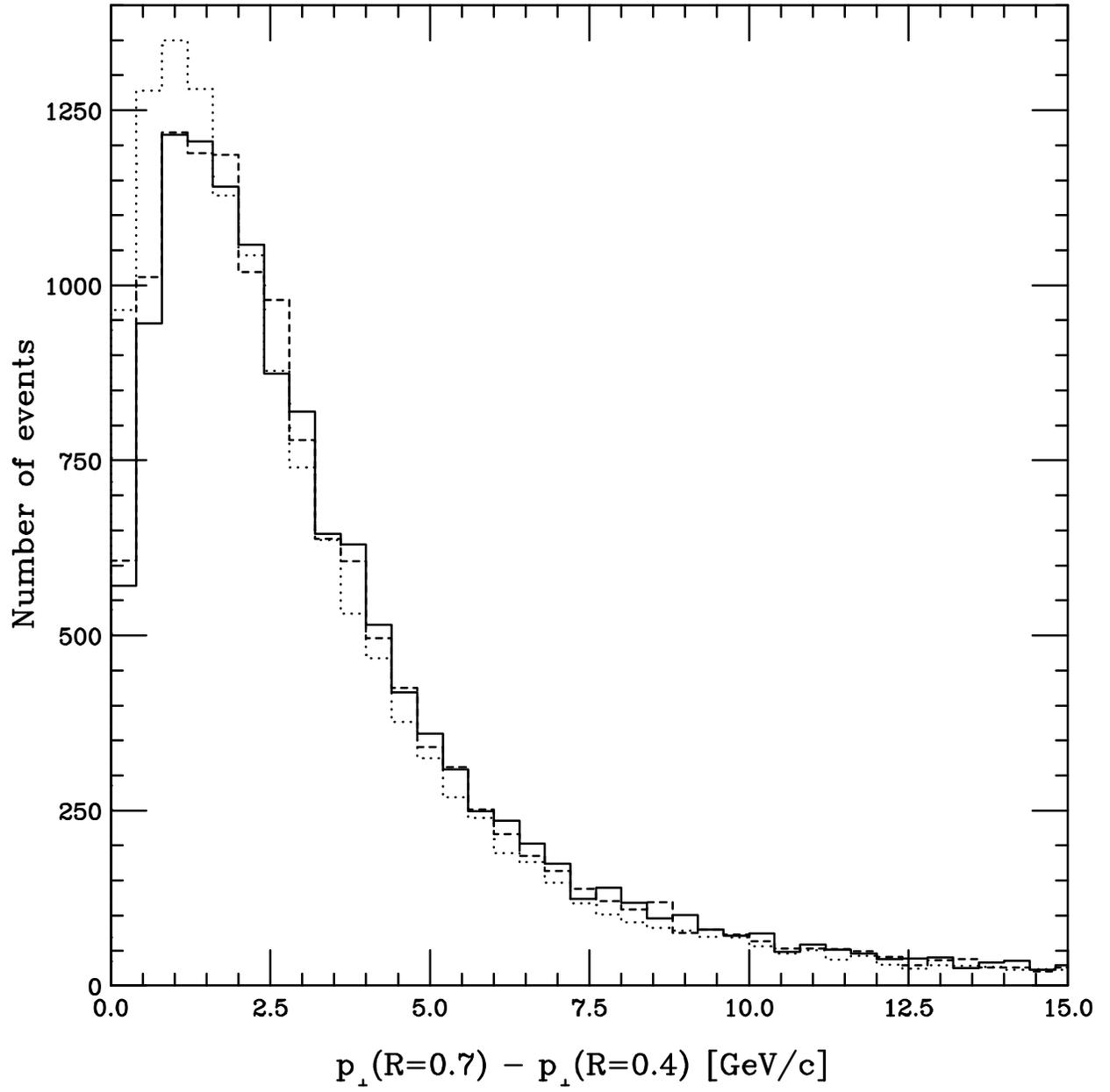
\begin{figure}
\caption{
The increase in jet $p_\perp$ when the cone
size is increased from $R=0.4$ to $R=0.7$ for jets that
contribute to the signal peak $155 < m_{\rm ave} < 185$,
with $p_\perp \approx 40$ in the smaller cones.
{\it Solid} curve = $b$-jets from leptonic top decay,
{\it dashed} curve = $b$-jets from hadronic top decay, and
{\it dotted} curve = jets from hadronic $W$ decay are all
about the same.  The large range of the distribution
directly shows the importance of using the larger cones to
analyze each jet, rather than attempting to make do with
smaller cones and compensating by an average
``out-of-cone correction.''}
\label{figure4}
\end{figure}

\end{document}